\documentclass[twocolumn,showpacs,a4,prl]{revtex4} 
\usepackage{graphicx}%
\usepackage{dcolumn} 
\usepackage{bm} 
\makeatletter \def\btt#1{\texttt{\@backslashchar#1}}%
\DeclareRobustCommand\bblash{\btt{\@backslashchar}}%
\makeatother  
\begin{document}  
\preprint{LCMO20.TEX}  
\title{Orbital domain state and finite size scaling in ferromagnetic insulating manganites}
\author{G. Papavassiliou$^1$, M. Belesi$^1$, M. Fardis$^1$, M. Pissas$^1$, J. Dolinsek$^2$, C. Dimitropoulos$^{1,\star }$, J. P. Ansermet$^{3,\star }$} 
\affiliation{$^1$Institute of Materials Science, NCSR,  Demokritos, 153 10 Aghia Paraskevi, Athens, Greece\\
$^2$Josef Stefan Institute, Jamova 39, 61111 Ljubljana, Slovenia\\
$^3$Dept. of Physics, University of Illinois, Urbana, Illinois 1801, USA} 
\altaffiliation{$^\star $Institut de Physique Exprimentale, EPFL-PH-Ecublens, 1015-Lausanne, Switzerland} 
\date{\today }   
\begin{abstract} $^{55}$Mn and $^{139}$La NMR measurements on a high quality single crystal of ferromagnetic (FM) La$_{0.80}$Ca$_{0.20}$MnO$_3$ demonstrate the formation of localized Mn$^{3+,4+}$ states below $70$ K, accompanied with strong anomalous increase of certain FM neutron Bragg peaks. $^{55,139}(1/T_1)$ spin-lattice relaxation rates diverge on approaching this temperature from below, signalling a genuine phase transition at $T_{tr}\simeq 70$ K.  The increased local magnetic anisotropy of the low temperature phase, the cooling-rate dependence of the Bragg peaks, and the observed finite size scaling of $T_{tr}$ with Ca (hole) doping, are suggestive of freezing into an orbital domain state, precursor to a phase transition into an inhomogeneous orbitally ordered state embodying hole-rich walls. 
\end{abstract} 
\pacs{75.70.Pa., 76.20.+q, 75.30.Et, 75.60.Ch} 
\maketitle  

Understanding the electronic properties of colossal magnetoresistive manganites has been a challenging subject for both experimentalists and theorists, ever since their discovery almost $50$ years ago. There are clearly two types of dominant ground states in these compounds: In La$_{1-x}$Ca$_x$MnO$_3$ (LCMO) for example, the ground state is ferromagnetic and metallic (FMM) for $0.2\leq x\leq 0.5$, and antiferromagnetic insulating for $x\geq 0.5$. The establishment of the FMM phase was initially attributed to the double exchange (DE) interaction \cite{Zener51}, i.e. ferromagnetism via the strong Hund's coupling between hopping $e_g$ electrons at neighbouring Mn$^{4+,3+}$ sites. However, the detection of FM insulating (FMI) \cite{Endoh99} and AFM metallic \cite{Kajimoto99} phases in certain manganites indicates that DE is inadequate for the full description of the magnetic and transport properties in these systems. According to recent theoretical \cite{Tokura00, Dagotto01,Mizokawa00} and experimental results \cite{Endoh99,Yamada96}, orbital ordering ($OO$) is an important factor controling the e$_g$-hole mobility. A characteristic example is La$_{1-x}$Sr$_x$MnO$_3$ (LSMO), where in the doping range $0.1\leq x\leq 0.15$ a FMM to FMI transition takes place at low temperatures \cite{Liu01,Zhou00}. Experiments have shown that this transition is associated with charge ordering, $OO$, and strong reduction of the cooperative Jahn-Teller (JT) lattice distortions in the low-T phase \cite{Endoh99,Yamada96,Yamada00}. It has been also proposed that the $OO$ phase might contain hole-rich layers \cite{Yamada96,Inami99}, which sets the question of stripe formation into the FMI phase \cite{Hotta01,Mizokawa00}.
 
A similar transition has been observed in LCMO for $0.125\leq x\leq 0.2$, at $T_{tr}\approx 70-100$ K \cite{Biotteau01,Papavassiliou00}. However, the characteristic resistivity upturn \cite{Okuda00,Markovich02a,Markovich02b}, which marks the onset of the FMI phase is observed at temperatures sufficiently higher than $T_{tr}$. Experiments show that the resistivity upturn is associated with a diffuse structural transition, characterised by strong reduction of the orthorhombicity \cite{Biotteau01}, and a remarkable rotation of the easy magnetization axis \cite{Markovich02b}. These characteristics are considered as the hallmark of orbital rearrangements that take place on cooling. At the same time, a number of peculiar features are observed, which are reminiscent of glassy freezing \cite{Dai00} especially in the doping region $0.17\leq x\leq 0.2$: (i) a steep decrease and frequency dependence of the ac susceptibility at low temperatures \cite{Markovich02a,Markovich02b,Laiho01}, (ii) strong difference between the field cooled (FC) and zero field cooled (ZFC) magnetization in low fields \cite{Markovich02b,Laiho01,Hong01,Belesi01}, (iii) the wipe-out of the NMR signal, which has been attributed to ultra-slow fluctuations of the electronic spin, charge, or orbital degrees of freedom \cite{Papavassiliou00,Belesi01,Papavassiliou01,Allodi01,Savosta02}. On the other hand, the sharp, cooling-rate dependent increase of the FM Bragg peaks \cite{Biotteau01} below $T_{tr}$, and the sudden slope-change in both the ZFC and FC branches of the magnetization at the same temperature \cite{Papavassiliou00,Markovich02b,Belesi01}, indicate rather nonequilibrium phenomena and quasinonergodicity ("freezing") on cooling, than a reentrant spin-glass transition. It is the competition between critical slowing down and spin freezing that makes characterization of the spin and orbital dynamics in this system a nontrivial task. 

In this paper we shed light on this intringuing freezing mechanism by using $^{139}$La and $^{55}$Mn NMR in comparison with recent neutron scattering measurements, performed on the same high quality single crystal of La$_{0.8}$Ca$_{0.2}$MnO$_3$. This system exhibits a PM-to-FM transition at $T_c\approx 180$ K, and resistivity upturn at $\approx 150$ K. We provide clear evidence about a novel thermodynamic phase transition occuring at $T_{tr}\approx 70$ K. The high magnetic anisotropy of the low temperature FMI phase and the finite size scaling of $T_{tr}$ upon increasing hole-doping, are indicative of an inhomogeneous $OO$ state with hole-stripes below $70$ K, while the spin-freezing features may arise from the formation of an indermediate FMI orbital domain state at nanometer length scale. This state creates partial magnetic disorder at orbital domain boundaries, while the establishment of long range $OO$ below $T_{tr}$ gives rise to full magnetic ordering, as deduced from the neutron scattering experiments  \cite{Biotteau01}. 

Zero external field $^{139}$La and $^{55}$Mn NMR measurements were acquired by applying a two pulse spin-echo technique, at very low rf power level, due to the very strong rf enhancement that characterizes FM materials \cite {Papavassiliou97}. $T_1$ was then measured at the peak of the spectra, by applying a saturation recovery technique and by fitting with a multiexponential recovery law as in previous works \cite{Papavassiliou01}. Rf enhancement experiments were performed by recording the NMR signal intensity $I$ as a function of the level of the applied rf field $H_1$. In general, the obtained curves follow an asymmetric bell-shaped law with maximum at $n\gamma H_1\tau =2\pi /3$, which allows the calculation of the rf enhancement factor $n$ \cite{Papavassiliou97}. The neutron scattering data have been recently reported in ref. \cite{Biotteau01}. 

\begin{figure}[tbp] 
\centering 
\includegraphics[angle=0,width=7cm]{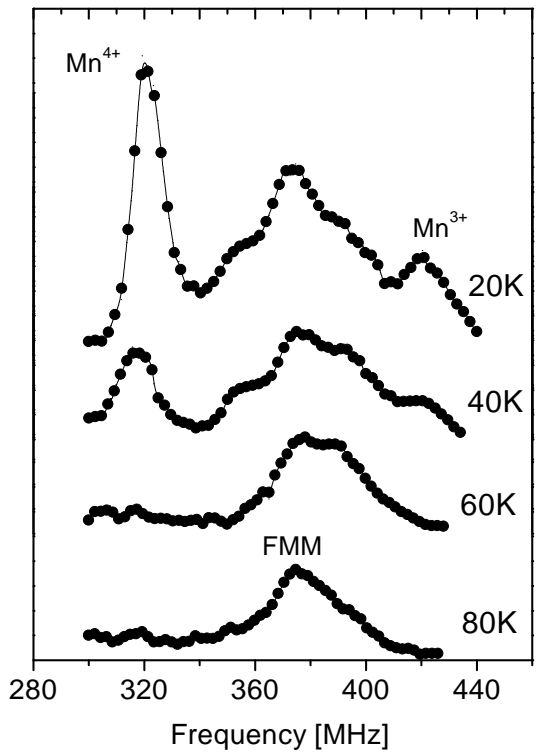} 
\caption{$^{55}$Mn NMR spectra of LCMO $x=0.20$, at various temperatures.}
\label{fig1} 
\end{figure}  

\begin{figure}[tbp] 
\centering 
\includegraphics[angle=0,width=7cm]{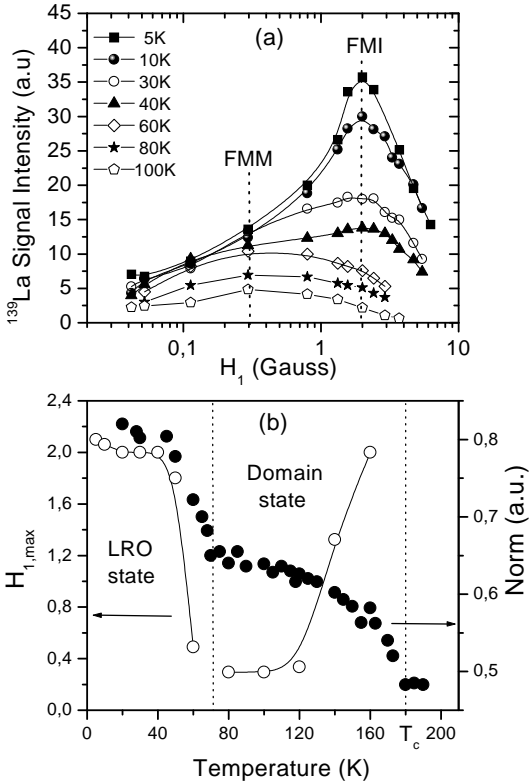} 
\caption{(a) $^{139}$La NMR signal intensity as a function of the rf field H$_1$ for LCMO $x=0.20$,  at various temperatures. (b) The rf field H$_{1,max}$ of maximum signal intensity as a function of temperature (o), together with the integrated intensity of the $(110)$ (or $(002)$) Bragg peaks vs. temperature ($\bullet$ ), from ref. \cite{Biotteau01}.}
\label{fig2} 
\end{figure}

Figure \ref{fig1} exhibits $^{55}$Mn line shape measurements at various temperatures. The maximum signal was obtained for $H_1\parallel$ c-axis, which means that the hyperfine field $H_{hf}$ is lying on the $ab$ plane. A similar result was obtained in $^{139}$La NMR in agreement with previous experiments on low doped LSMO \cite{Kumagai99}. For $T\geq 70$K spectra consist of a broad single-peaked line at frequency $\approx 375$ MHz, which according to the literature corresponds to delocalized Mn states \cite{Papavassiliou00}). However, for $T\leq 70$K new peaks increase rapidly on cooling, which correspond to localized Mn$^{4+}$ states (the narrow peak at $\approx 320$ MHz) and Mn$^{3+}$ states (the broad peak at $\approx 420$ MHz) \cite{Papavassiliou00}). At exactly the same temperature a sharp increase of the intensity of certain ferromagnetic Bragg peaks (Figure \ref{fig2}b) is observed on cooling, which implies that the appearance of the localized Mn$^{3+,4+}$ NMR peaks is associated with better ordering of the Mn spins. The intensity of these Bragg peaks depends on the cooling rate as recently reported in ref. \cite{Biotteau01}. The rapid increase of the NMR signal from localized electron states below $70$ K is also clearly deduced from the $^{139}$La rf enhancement experiments of Figure \ref{fig2}a. For $T\geq 70$ K only a broad peak with maximum at $0.3$ Gauss is present in the $I$ vs. $H_1$ curves. However, below $70$ K a second peak at $\approx 2$ Gauss appears, which increases rapidly by decreasing temperature. Comparison with Figure \ref{fig1} implies that this second peak corresponds to localized Mn$^{3+,4+}$ states, whereas the $0.3$ Gauss peak to delocalized Mn states. There are two possible explanations for the different rf enhancement factors of the FMI and FMM signals: (i) the $OO$ FMI matrix phase has higher local magnetic anisotropy (smaller response to the applied rf power level) than the orbitally disorder FMM minority phase \cite{Belesi01}, (ii) the system consists of hole-rich FMM domain-walls separating FMI domains. The second argument is based on the fact that the rf enhancement $n\approx 3\cdot 10^3$ from the FMM regions is sufficiently higher than the FMI $n\approx 2\cdot 10^2$, as expected for FM domain-walls in comparison to FM domains \cite{Papavassiliou97}. Finite size scaling arguments given below, are in favour of the second explanation.  

\begin{figure}[tbp] 
\centering 
\includegraphics[angle=0,width=7cm]{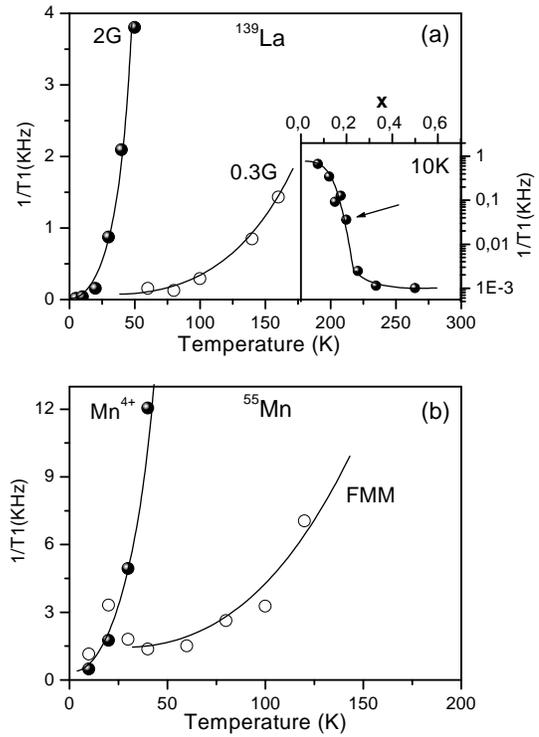} 
\caption{(a)$^{139}(1/T_1)$ of LCMO $x=0.20$, as a function of temperature. The inset shows  $^{139}(1/T_1)$ of powder LCMO samples, as a function of Ca doping at $T=10$ K and rf level $2$ Gauss. The arrow indicates the value for the $x=0.20$ single crystal. (b)$^{55}(1/T_1)$ of La$_{0.80}$Ca$_{0.20}$MnO$_3$ as a function of temperature.}
\label{fig3} 
\end{figure}

Figure \ref{fig3} shows $^{55,139}(1/T_1)$ measurements as a function of temperature. In case of $^{55}(1/T_1)$ measurements were performed on the M$^{4+}$ peak and the central FMM peak, whereas in case of $^{139}(1/T_1)$ measurements were obtained at rf power levels $0.3$ Gauss and $2$ Gauss, which correspond to delocalized and localized Mn$^{3+,4+}$ states, respectively. The similar temperature dependence  of the corresponding $^{55,139}1/T_1$ curves is a proof of the correct assignement of the $^{139}$La NMR signals. The crucial point in Figures \ref{fig3}a,b is that both $^{55,139}1/T_1$ from Mn$^{3+,4+}$ ions are diverging on approaching $70$K from below. Such a behaviour is indicative of critical relaxation enhancement on appoaching the phase transition temperature. In a previous work on a series of powder samples the divergence of $1/T_1$ was not observed \cite{Papavassiliou01}, probably due to inaccurate irradiation conditions in non-oriented powder samples. In contrast, the FMM $1/T_1$ is insensitive to the phase transition while crossing $T_{tr}$. The hump at $\approx 20$K on FMM $^{55}(1/T_1)$ is probably due to partial signal contribution at $\approx 375$ MHz from localized Mn$^{3+}$ states, which induce a broad inhomogeneous signal extending down to $\approx 350$ MHz \cite{Papavassiliou00,Anane96}. At temperatures higher than $70$K, where the FMI signal is completely wiped out \cite{Papavassiliou01,Allodi01}, $1/T_1$ reflects solely the dynamics of the FMM states. 

\begin{figure}[tbp] 
\centering 
\includegraphics[angle=0,width=7cm]{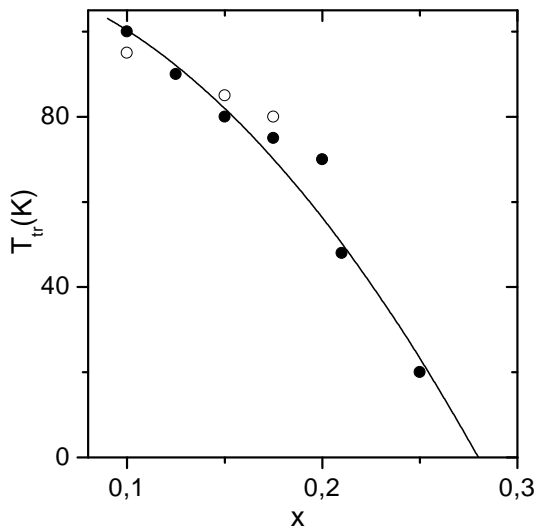} 
\caption{           $T_{tr}$ for LCMO vs. doping $x$. Experimental points were obtained by NMR ($\bullet$) and magnetic measurements ($\circ$). The solid line is theoretical fit as described in the text.}
\label{fig4} 
\end{figure}

The inset in Figure \ref{fig3}a shows $^{139}(1/T_1)$ as a function of doping obtained in powder LCMO samples at $T=10$ K. It is observed that in the pure FMM part of the phase diagram ($x\geq 0.25$), $1/T_1$ is almost three orders of magnitude lower than in the pure FMI part ($x\leq 0.15$). It is thus remarkable that for $x=0.20$ no clear distinction in the $1/T_1$ values from FMI and FMM states is observed at low temperatures, while $1/T_1$'s become particularly different on approaching $T_{tr}$ from below. In a recent neutron scattering study (performed on the same $x=0.20$ crystal as here) \cite{Hennion02} two different FM media were detected, which coexist dynamically in the temperature range $T_{tr}\leq T\leq T_c$, and are stabilized into a periodically arranged collective state below $T_{tr}$. Hence, at low temperatures the system appears to consist of a regular arrangement of $OO$ domains with a uniform relaxation mechanism. In order to envisage how such an orbital domain state could be created, we consider the idea of the random field Ising model \cite{Fishman79,Bruinsma84}, where the random field mimics the lattice distortions induced by substitution with Ca ions. By taking into account only the isospin $T =1/2$ degree of freedom, which describes the twofold $e_g$ orbital degeneracy \cite{Tokura00, Dagotto01}, it can be shown that in strong random fields an intermediate orbital domain state may be realized on cooling \cite{Ro85,Wong83,Yoshizawa84}. This state is metastable exhibiting anomalous slow relaxation with logarithmic time dependence \cite{Nattermann88,Villain84} and strong difference in the orbital and spin ordering between the ZFC and FC branches \cite{Ro85}. The observed cooling rate dependence of the FM Bragg peak intensity below $T_{tr}$  is attributed to nucleation or rearrangement of orbital domains and domain walls, which is connected with the large reduction of orthorhombicity upon cooling \cite{Biotteau01}.  
We also note that in case that neighboring orbital domains are arranged in antiphase, stripe-like orbital walls will be formed, where holes are energetically favorable to concentrate \cite{Hotta01}, in agreement with the neutron measurements \cite{Hennion02}. By increasing hole-doping the number of walls will increase \cite{Hotta01}, whereas above a critical doping $x_c$ the $OO$ phase is expected to be supressed \cite{Fisher72,Cho93}. Indeed, such doping induced finite size scaling effects are experimentally suggested. Figure \ref{fig4} shows $T_{tr}$ vs. doping $x$, obtained from NMR and magnetic measurements in a series of LCMO powder samples. The experimental data for $0.1\leq x\leq 0.25$ are nicely fitted by the expression $T_{tr}(x)=T^*_{tr}(1-(x/x_c)^n)$, where $T^*_{tr}=110$ K, $x_c=0.28$, and $n=2$. Such a power-law dependence is expected from finite size scaling theory \cite{Fisher72,Cho93}, which predicts that the effective $T_{tr}$ is limited by the finite size $L$ of the domains according to the formula, $T_{tr}(L)=T_{tr}(\infty)(1-(L)^{-1/\nu})$. The above equations are consistent with $L(x)=1/x^{n\nu }$, which is determined as evidence about the formation of walls separating orbital domains. We also notice that in mean-field approximation $\nu =1/2$, so that $L(x)=1/x$, which indicates that the wall width is independent of $x$ in accordance with theoretical predictions \cite{Hotta01}. 

In summary, NMR in conjunction with recent neutron scattering experiments on a single crystalline LCMO $x=0.20$, indicate the presence of a novel phase transition from a high temperature orbital domain state to a low temperature state of arranged orbital domains. Although no direct evidence for stripe formation is claimed here, the difference in the rf enhancement between the FMM and the FMI phases, and the finite size scaling effects on the transition temperature T$_{tr}$, suggest that the formation of the domain state and the orbital and spin freezing effects are possibly associated with the formation of hole stripes. By assigning the $e_g$ orbital degree of freedom to an isospin, there is a complete analogy between the $OO$ phase in manganites and the low temperature hole-striped phase in cuprates and nickelates \cite{Cho93,Tranquada95}. This is probably due to the similar competitive spin (isospin) ordering under quenched disorder \cite{Burgy01} in the layered (2D) structures. We notice the impressive similarity between our NMR results in  LCMO $x=0.20$, and the $^{139}$La NMR in the stripe-ordered nickelate La$_{5/3}$Sr$_{1/3}$NiO$_4$ \cite{Yoshinari99}. Clearly, further investigations of the orbital and spin dynamics in lightly doped manganites will be particularly helpful in understanding the way that holes are self-assembled - apparently in stripes - in these systems.   

This work has been partially supported by the Greek-Slovenian cooperation project No. 2495. We would like to thank Dr. M. Hennion, and Prof. Y. M. Mukovskii, for providing the $x=0.20$ single crystal.

\end{document}